\documentclass[aps,pra,reprint,twocolumn]{revtex4}
\usepackage[dvipdfmx]{graphicx}
\usepackage{graphicx}
\usepackage{color}
\usepackage{amsmath,amssymb}
\usepackage{bm}
\usepackage{comment}
\usepackage{ulem}

\begin{document}


\title{Decay mechanisms of superflow of Bose-Einstein condensates in ring traps}


\author{Masaya Kunimi}
\email{E-mail: masaya.kunimi@yukawa.kyoto-u.ac.jp}
\author{Ippei Danshita}
\thanks{Present address : Department of Physics, Kindai University, Higashi-Osaka, Osaka 577-8502, Japan}
\affiliation{Yukawa Institute for Theoretical Physics, Kyoto University, Kyoto 606-8502, Japan}



\date{\today}

\begin{abstract}
We study the supercurrent decay of a Bose-Einstein condensate in a ring trap combined with a repulsive barrier potential. In recent experiments, Kumar {\it et al.} [Phys. Rev. A {\bf 95}, 021602(R) (2017)] have measured the dependence of the decay rate on the temperature and the barrier strength. However, the origin of the decay observed in the experiment remains unclear. We calculate the rate of supercurrent decay due to thermally activated phase slips (TAPS) by using the Kramers formula based on the Gross-Pitaevskii mean-field theory. The resulting decay rate is astronomically small compared to that measured in the experiment, thus excluding the possibility of TAPS as the decay mechanism. Alternatively, we argue that three-body losses can be relevant to the observed decay and predict that one can observe supercurrent decay via TAPS by decreasing the number of atoms.
\end{abstract}

\maketitle

\section{Introduction}\label{sec:Introduction}

Systems of ultracold gases confined in ring-shaped traps have served as an ideal platform for studies of superfluidity~\cite{Ryu2007,Ramanathan2011,Moulder2012,Beattie2013,Neely2013,Wright2013,Wright2013_2,Ryu2013,Ryu2014,Eckel2014,Jendrzejewski2014,Eckel2014_2,Corman2014,Kumar2017}. Their exquisite controllability has led to experimental observations of various fundamental properties of superfluids, including persistent currents \cite{Ramanathan2011,Moulder2012,Beattie2013,Eckel2014_2,Corman2014,Kumar2017}, critical velocities \cite{Wright2013,Wright2013_2}, and hysteresis loops \cite{Eckel2014}. Such systems attract growing interest also as an atomic analog of superconducting quantum interference device \cite{Clarke2004}, which constitutes a basic element for atomtronic circuits \cite{Seaman2007,Pepino2009}. 

\begin{figure}[t]
\centering
\includegraphics[bb=0 0 350 250,width=7.5cm,clip]{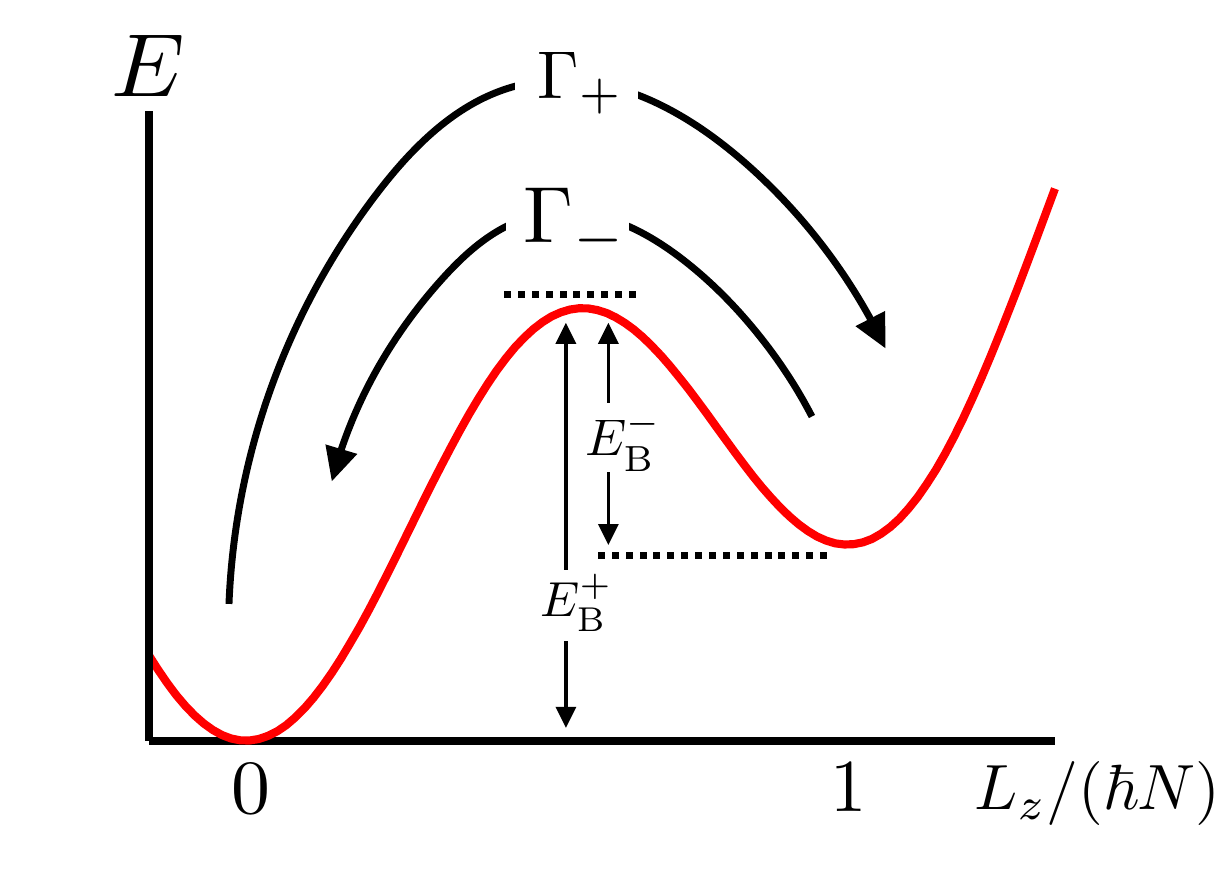}
\caption{(Color online) Schematic picture of the energy landscape of the system. The left and right local minima are the ground state and a persistent current state, respectively. The local maximum is an unstable state, in the present case, the solitonic vortex state. $\Gamma_{\pm}$ and $E_{\rm B}^{\pm}$ are the nucleation rate and the energy barrier of the acceleration and the decay process, respectively.}
\label{fig:energy_landscape}
\end{figure}

A recent experiment performed by the NIST group has raised a puzzling question concerning the superfluidity in a ring-shaped Bose-Einstein condensate (BEC)~\cite{Kumar2017}. In this experiment, they prepared a current-carrying BEC with winding number 1 as an initial state and measured the lifetime of this state. To induce the decay of the persistent current, they exposed the BEC to a repulsive potential barrier, which is generated by a blue detuned laser beam. They found that the observed decay rate is significantly dependent on the barrier strength $U_0$ and the temperature $T$. One might naively expect that this result could be interpreted as a decay of the persistent current via thermally activated phase slips (TAPS)~\cite{Langer_Fisher1967,Langer1967,McCumber1970}, whose decay rate $\Gamma$ obeys the Arrhenius law, i.e., $\Gamma \propto e^{-E_{\rm B}^{-}/(k_{\rm B}T)}$. Here $E_{\rm B}^{-}$ denotes the energy barrier separating the state with winding number 1 from that with winding number zero (see Fig.~\ref{fig:energy_landscape}).  However, the NIST group presented a rough estimation of $E_{\rm B}^{-}$ to argue that the observed decay rate is inconsistent with the Arrhenius law. Thus the decay mechanism of the persistent currents remains unclear.

The above-mentioned observation of Ref.~\cite{Kumar2017} challenges our common understanding of the superfluidity in the sense that TAPS has been established as a universal decay mechanism applicable to various superfluid systems at finite temperatures, such as superfluid ${}^4$He \cite{Langer1970,Reppy1992}, superconductors \cite{Halperin2010}, spin superfluids \cite{Kim2016}, and one-dimensional Bose gases in optical lattices \cite{Tanzi2016,Kunimi2017}. Resolving this puzzle is important also for engineering and controlling the atomtronic circuits, which require quantitative understanding of the persistent current in the presence of repulsive potential barriers~\cite{Ryu2013}.

In this paper, we first reexamine the possibility of TAPS as a decay mechanism by quantitatively computing the nucleation rate of TAPS without any fitting parameters. Our quantitative result shows that the energy barrier in Ref.~\cite{Kumar2017} was considerably underestimated. Nevertheless, the computed decay rate completely disagrees with the value measured in Ref.~\cite{Kumar2017}, thus confirming that TAPS is irrelevant to the observed decay. Alternatively, we find that three-body losses can induce the decay of superflow in the experimental setup and discuss the possibility of this decay mechanism in the experiment. This indicates that one may observe decay via TAPS by optimizing some experimental parameters in such a way that TAPS is enhanced. We explicitly identify parameter regions in which TAPS dominates over three-body losses and predict the decay rate via TAPS as a reference that can be directly compared with future experiments.

This paper is organized as follows. In Sec.~\ref{sec:model}, we explain the system considered in this paper. In Secs.~\ref{subsec:Quasi-2D} and \ref{subsec:3D}, we show our results for the energy barrier, lifetime, and real-time dynamics within the three-body loss in the quasi-2D and 3D systems. In Sec.~\ref{sec:Summary}, we summarize our results. In the Appendix, we explain the details of the numerical method used in this paper.

\section{Model and methods}\label{sec:model}
We consider a BEC in a combined potential of a ring trap and a repulsive barrier to mimic the situation in the experiment of Ref.~\cite{Kumar2017}. The interatomic interaction is sufficiently weak so that the dynamics of the superfluid order parameter $\psi(x, y, z, t)$ is quantitatively described within the mean-field approximation. Specifically, the three-dimensional (3D) Gross-Pitaevskii (GP) equation with a phenomenological three-body loss term \cite{Kagan1998} is given by
\begin{align}
&i\hbar\frac{\partial}{\partial t}\psi(x, y, z, t)\notag \\
&=\left[-\frac{\hbar^2}{2m}\bm{\nabla}^2+U(x,y,z)+g|\psi(x, y, z, t)|^2\right]\psi(x, y, z, t)\notag \\
&\quad -\frac{i\hbar}{2}L_{3}|\psi(x, y, z, t)|^4\psi(x, y, z, t),\label{eq:3D_GP_equation}
\end{align}
where $m$ is the mass of the atom, $U(x, y, z)\equiv (1/2)m\omega_{\rm r}^2(r-R)^2+(1/2)m\omega_z^2z^2+U_{\rm ext}(\bm{r})$ represents the trap potential and the barrier potential, ${\bm r} = (x,y)$, $\omega_{\rm r}$ and $\omega_z$ are the trap frequency of the radial direction and $z$ direction, respectively, $R$ is the mean radius of the ring trap, $g\equiv 4\pi\hbar^2a_{\rm s}/m$ is the coupling constant, $a_{\rm s}$ is the s-wave scattering length, and $L_3$ is the three-body loss rate. We note that the three-body loss term is set to zero except for the calculations of Fig.~\ref{fig:angular_momentum_3D} in Sec.~\ref{sec:Three-body_loss_3D}. The barrier potential is created by dithering the Gaussian laser beam. The time-averaged potential $U_{\rm ext}(\bm{r})$ is given by \cite{Kumar2017_pb}
\begin{align}
U_{\rm ext}(\bm{r})&=\frac{U_0}{2}\left\{{\rm erf}\left[\frac{\sqrt{2}}{w}\left(x-R+\frac{l_{\rm d}}{2}\right)\right]\right.\notag \\
&\hspace{3.5em}\left.-{\rm erf}\left[\frac{\sqrt{2}}{w}\left(x-R-\frac{l_{\rm d}}{2}\right)\right]\right\}e^{-2y^2/w^2},\label{eq:external_potential}
\end{align}
where $U_0$ is the potential strength, $w$ is the $1/e^2$ width of the laser beam, $l_{\rm d}$ is the length of the dither, and ${\rm erf}(\cdot)$ is the error function. 

In Sec.~\ref{subsec:Quasi-2D}, we show the quasi-two-dimensional (2D) results. The quasi-2D GP equation is derived by substituting $\psi(x, y, z, t)=\Psi(\bm{r}, t)[1/\sqrt{\sqrt{\pi} a_z}]e^{-(z/a_z)^2/2}$ into Eq.~(\ref{eq:3D_GP_equation}):
\begin{align}
&i\hbar\frac{\partial}{\partial t}\Psi(\bm{r}, t)\notag \\
&=\left[-\frac{\hbar^2}{2m}\bm{\nabla}_{\rm 2D}^2+U_{\rm 2D}(\bm{r})+g_{\rm 2D}|\Psi(\bm{r}, t)|^2\right]\Psi(\bm{r}, t)\notag \\
&\quad-\frac{i\hbar}{2}L_{{\rm 3, 2D}}|\Psi(\bm{r}, t)|^4\Psi(\bm{r}, t),\label{eq:quasi-2D_GP_eq}
\end{align}
where $\bm{\nabla}^2_{\rm 2D}\equiv \partial^2/\partial x^2+\partial^2/\partial y^2$ is the two-dimensional Laplacian, $U_{\rm 2D}(\bm{r})\equiv (1/2)m\omega_{\rm r}^2(r-R)^2+U_{\rm ext}(\bm{r})$ is the two-dimensional external potential, $g_{\rm 2D}\equiv \sqrt{8\pi}\hbar^2a_{\rm s}/(m a_z)$ is the two-dimensional coupling constant \cite{Petrov2000_2},  $a_{z}\equiv \sqrt{\hbar/(m\omega_z)}$ is the harmonic-oscillator length in the $z$ direction, and $L_{3,{\rm 2D}}\equiv L_3/(\sqrt{3}\pi a_z^2)$ is the quasi-2D three-body loss rate. We also note that the $L_{{\rm 3,2D}}$ term is set to zero except for the calculations of Fig.~\ref{fig:angular_momentum}. in Sec.~\ref{sec:Three-body_loss_2D}.

To obtain excitation spectra, we linearize the GP equation around a stationary solution $\Psi(\bm{r}, t)=e^{-i \mu t/\hbar}\Phi(\bm{r})$, where $\mu$ is the chemical potential. The linearized GP equation, i.e., the Bogoliubov equation, is given by
\begin{align}
H_{\rm B}
\begin{bmatrix}
u_i(\bm{r}) \\
v_i(\bm{r})
\end{bmatrix}
&=\epsilon_i
\begin{bmatrix}
u_i(\bm{r}) \\
v_i(\bm{r})
\end{bmatrix}
,\label{eq:Bogoliubov_equation}\\
H_{\rm B}&\equiv 
\begin{bmatrix}
\mathcal{L} & g_{\rm 2D}\Phi(\bm{r})^2 \\
-g_{\rm 2D}\Phi^{\ast}(\bm{r})^2 & -\mathcal{L}^{\ast}
\end{bmatrix}
,\label{eq:definition_of_HB}\\
\mathcal{L}&\equiv -\frac{\hbar^2}{2m}\bm{\nabla}^2_{\rm 2D}+U_{\rm 2D}(\bm{r})-\mu+2g_{\rm 2D}|\Phi(\bm{r})|^2,\label{eq:definition_of_L}
\end{align}
where $\epsilon_i$ is the excitation energy and $u_i(\bm{r})$ and $v_i(\bm{r})$ are eigenfunctions of the excited state labeled by $i$.

In addition to the Bogoliubov equation, we need to diagonalize the following matrix: $H_{\rm E}\equiv \sigma_3H_{\rm B}$, where $\sigma_3={\rm diag}(+1,-1)$. Let $\lambda_i$ be an eigenvalue of the matrix $H_{\rm E}$, which is a real value in contrast to the eigenvalue of the Bogoliubov equation $\epsilon_i\in\mathbb{C}$.

In the experiment of Ref.~\cite{Kumar2017}, they prepare a state with the winding number $W=1$ as the initial state. Here, we calculate the decay from $W=1$ state (metastable state) to $W=0$ state (ground state). According to the Kramers formula \cite{Langer1969,Hanggi1990}, the nucleation rate can be written as
\begin{align}
\Gamma&\equiv \Gamma_--\Gamma_+,\label{eq:definition_of_Gamma}\\
\Gamma_{\pm}&\equiv \frac{|\epsilon_{\rm DI}|/\hbar}{2\pi}\prod_n{}^{'}\sqrt{\frac{\lambda_n^{\rm s\pm}}{|\lambda_n^{\rm u}|}}e^{-E_{\rm B}^{\pm}/(k_{\rm B}T)},\label{eq:Kramers_formula}
\end{align}
where $\Gamma_- (\Gamma_+)$ is a nucleation rate of the decay process (the acceleration process) (see Fig.~\ref{fig:energy_landscape}), $\epsilon_{\rm DI}$ is the frequency of the unstable mode and $\lambda_n^{\rm u}$ and $\lambda_n^{\rm s\pm}$ are the eigenvalues of the matrix $H_{\rm E}$, $E_{\rm B}^{\pm}$ is the energy barrier of the acceleration and the decay process, $k_{\rm B}$ is the Boltzmann constant, and $T$ is the temperature of the system. In the product of Eq.~(\ref{eq:Kramers_formula}), we omit the zero modes ($\lambda_n=0$). We define the nucleation rate (inverse of the lifetime) as the difference between $\Gamma_-$ and $\Gamma_+$. We note that the Kramers formula is valid for $E_{\rm B}^{\pm}\gg k_{\rm B}T$.

Here, we explain the numerical method used in this work. In the quasi-2D calculations of the GP equation and the Bogoliubov equation, we use the almost same methods as those used in our previous work  \cite{Kunimi2017}, i.e., the space discretization is performed by the discrete variable representation method \cite{Baye1986} (see also the Appendix \ref{sec:Appendix_DVR}) and seeking the unstable stationary solution of the GP equation is based on the pseudoarclength continuation method \cite{Keller1987,Kunimi2015} and the Newton method. For the real-time dynamics, we use the pseudospectral method. The typical numerical meshes taken in this work are $129\times 129$. 

In the 3D calculation, we use the standard discretization method for the calculations of the energy barrier (see Sec.~\ref{subsec:3D}) and pseudospectral method for the calculations of the real-time dynamics. The typical numerical meshes taken in this work are $129\times 129\times 65$.

It is worth emphasizing that although the GP equation has been extensively used for studying superfluidity of ring-shaped BEC \cite{Brand2001,Piazza2009,Mathey2014,Mateo2015,Snizhko2016}, decay rates via TAPS at 2D, to our knowledge, have never been quantitatively calculated because of the difficulty in finding the unstable solutions.

\section{Results}\label{sec:Results}

\subsection{Quasi-2D results}\label{subsec:Quasi-2D}

\begin{figure}[t]
\centering
\includegraphics[width=8.5cm]{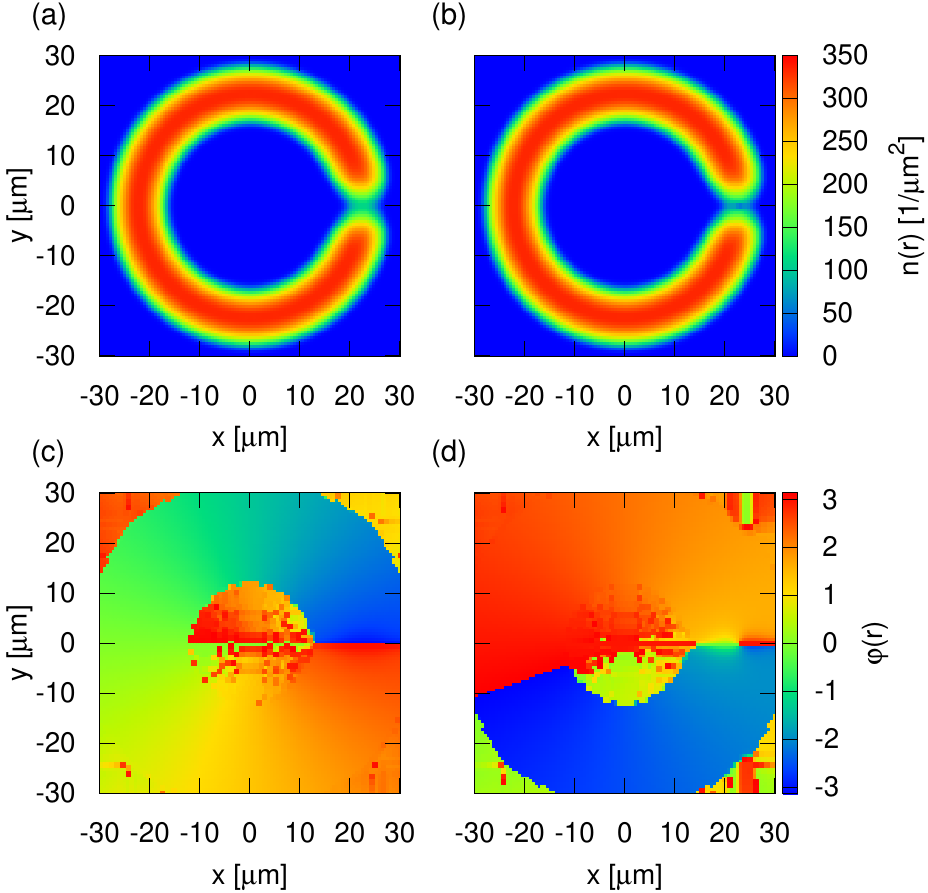}
\caption{(Color online) Density [$n(\bm{r})\equiv |\Phi(\bm{r})|^2$] and phase [$\varphi(\bm{r})\equiv {\rm Im}\log \Phi(\bm{r})$] profiles of the metastable [(a) and (c)] and unstable [(b) and (d)] solutions for $U_0=0.7\mu_{\rm exp}$, respectively.}
\label{fig:density_and_phase_profile}
\end{figure}%

Here, we show our theoretical results for quasi-2D systems, which are compared to the experimental ones. In the following calculations, we consider ${}^{23}$Na atom and set the system parameters as the case I in Refs.~\cite{Kumar2017,Supplemental_Kumar}, which corresponds to the lowest temperature and the strongest two dimensionality in the experiment. The specific values are as follows: $m=3.82\times 10^{-26}\;{\rm kg}$, $(\omega_{\rm r}, \omega_z)/(2\pi)=(258\;{\rm Hz}, 974\;{\rm Hz})$, $R=22.4\;\mu{\rm m}$, $a_{\rm s}=2.75\;{\rm nm}$, $w=6\;\mu{\rm m}$, and $l_{\rm d}=21.8\;\mu{\rm m}$. We use the particle number $N\simeq 3.75\times 10^5$, which comes from the ground-state particle number at the chemical potential $\mu=\mu_{\rm exp}\equiv h\times 2.91\;{\rm kHz}$. Here, $\mu_{\rm exp}$ is the chemical potential measured in the experiment.

\subsubsection{Density and phase profiles}\label{sec:density_and_phase_2D}

First, we show the density and phase profiles of the order parameter for the metastable and unstable states in Fig.~\ref{fig:density_and_phase_profile}. We see a solitonic vortex (SV) at the low-density region under the barrier potential in the unstable solution [Figs.~\ref{fig:density_and_phase_profile} (b) and \ref{fig:density_and_phase_profile} (d)]. The NIST group assumed the presence of such a SV in the unstable state when they estimated the energy barrier \cite{Kumar2017}. Our results support the validity of their assumption.

\subsubsection{Energy barrier and lifetime}\label{sec:energy_barrier_2D}

\begin{figure}[t]
\centering
\includegraphics[width=8.0cm]{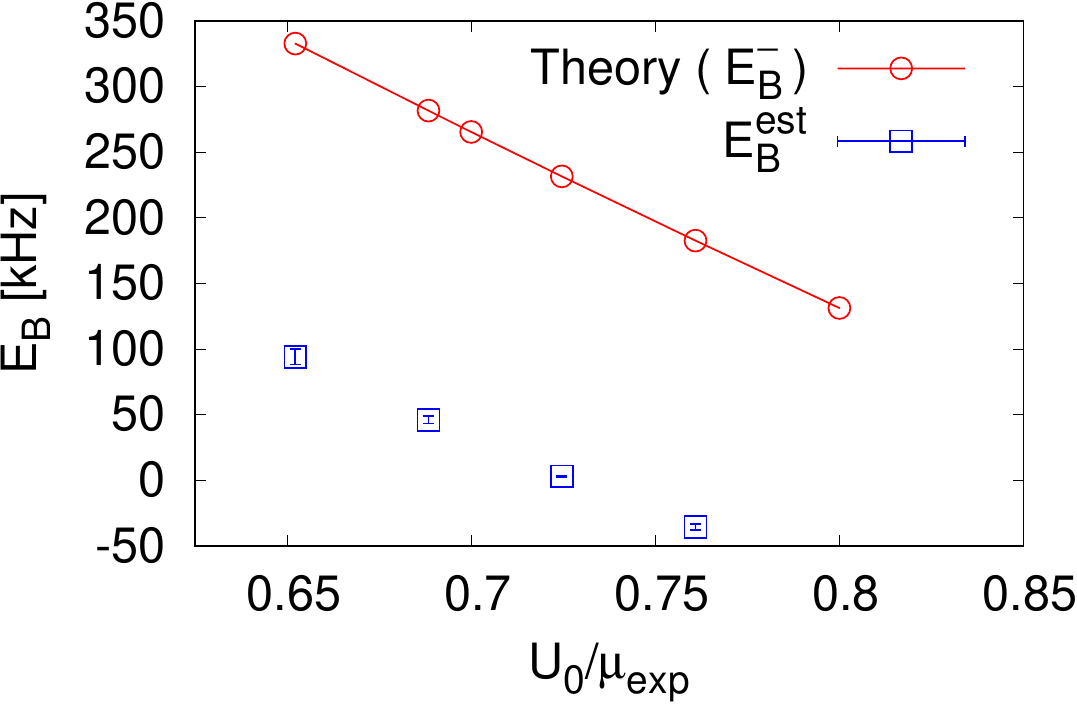}
\caption{(Color online) Energy barrier as a function of the strength of the external potential. The red circles and the blue squares denote the values computed in this work and those estimated in Ref.~\cite{Kumar2017}, respectively. 
}
\label{fig:energy_barrier}
\end{figure}

Next, we show the energy barrier as a function of the strength of the external potential in Fig.~\ref{fig:energy_barrier}. There we compare our results with those estimated in Ref.~\cite{Kumar2017} and see that the latter is considerably underestimated. This happens because the energy barrier was estimated by the energy of the SV [Eq.~(2) in Ref.~\cite{Kumar2017}] for a uniform system. The present system is nonuniform due to the trap potential and the external potential, which significantly modify the quantitative size of the energy barrier.

\begin{figure}[t]
\centering
\includegraphics[width=8.0cm]{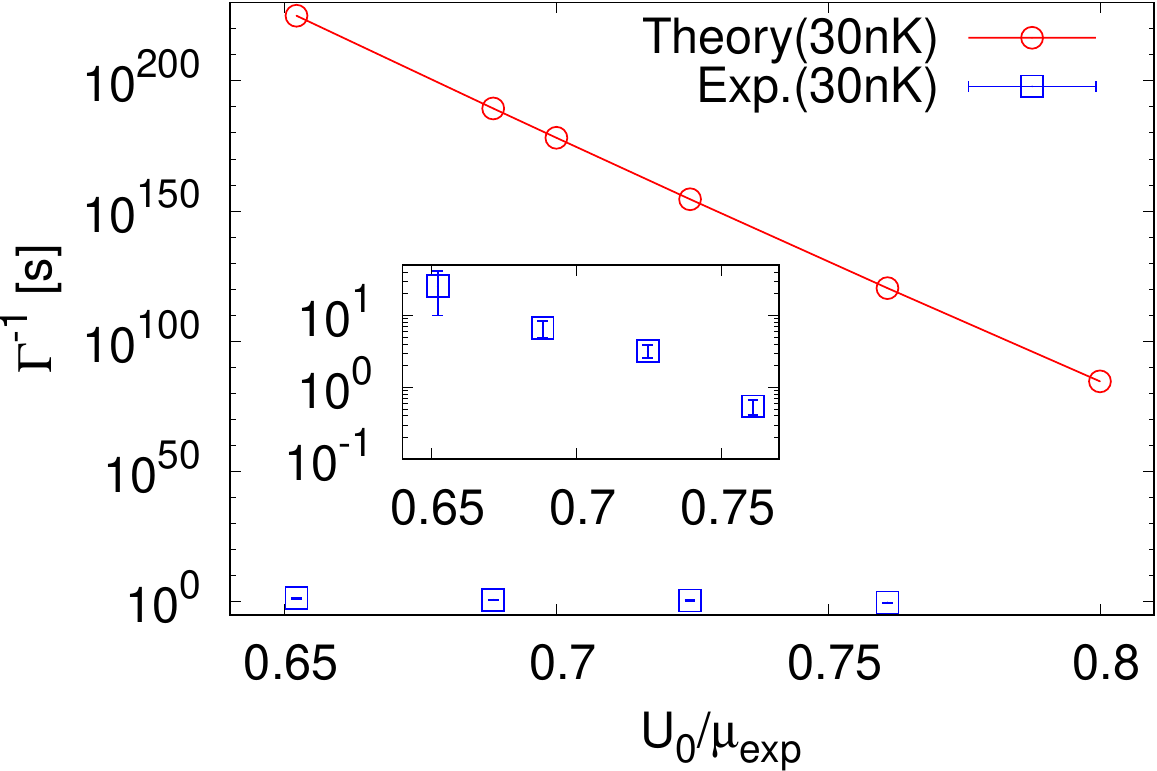}
\caption{(Color online) Lifetime of the superflow ($\Gamma^{-1}$) as a function of the strength of the external potential at $T=30\;{\rm nK}$. The red circle and the blue square denote the theoretical and the experimental results, respectively. The inset shows the magnified view of the experimental lifetime.
}
\label{fig:lifetime_vs_u}
\end{figure}%

We then show the lifetime of the superflow due to the TAPS in Fig.~\ref{fig:lifetime_vs_u}. We see that the lifetime due to TAPS is astronomically longer than that observed in Ref.~\cite{Kumar2017} ($1$-$10\;{\rm s}$). The origin of such long lifetime is the large energy barrier compared to the temperature. In the present case, the energy barrier is typically $E_{\rm B}/h\sim 200\;{\rm kHz}$ (see Fig.~\ref{fig:energy_barrier}) and the temperature is given by $k_{\rm B}T/h\simeq 0.6\;{\rm kHz}$ $(T=30\;{\rm nK})$. Consequently, we obtain the extremely long lifetime $\Gamma^{-1}\sim e^{E_{\rm B}/(k_{\rm B}T)}\sim e^{300}$. From the above results, we conclude that the TAPS is irrelevant to the decay of superflow observed in Ref.~\cite{Kumar2017}.

\subsubsection{Three-body loss}\label{sec:Three-body_loss_2D}

\begin{figure}[t]
\centering
\includegraphics[width=8.0cm]{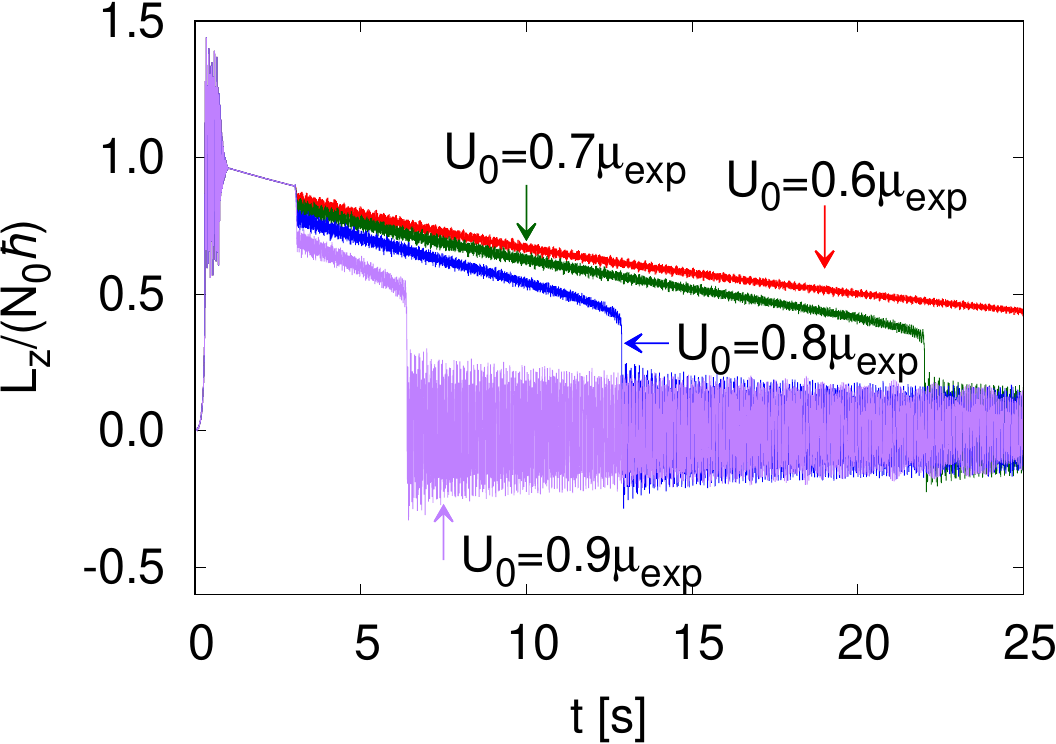}
\caption{(Color online) Time evolution of the angular momentum for various strength of the potential. The angular momentum is normalized by the initial particle number $N_0\simeq 4.40\times 10^5$.
}
\label{fig:angular_momentum}
\end{figure}%

Instead of the TAPS, we investigate the effects of three-body losses. The reason why we consider three-body losses is that the time scale of the NIST experiment ($\sim 5{\rm s}$) is longer than that of the typical cold atomic gas experiments ($\lesssim 1{\rm s}$). Hence the three-body loss, which is negligible in the usual cases, may become relevant to the supercurrent decay. Specifically, the superflow state is a metastable state, which may be broken by three-body losses.

In order to investigate the effects of three-body losses in quasi-2D systems, we numerically solve Eq.~(\ref{eq:quasi-2D_GP_eq}) with $L_{3,\rm{2D}}\not=0$. Here, we set $L_3=1.1\times 10^{-30}\;{\rm cm^{-6}/s}$, which is the three-body loss rate of the ${}^{23}$Na atom \cite{Stamper-Kurn1998}. The initial condition is the ground-state solution of the nondissipative GP equation ($L_3=0$) for $U_0=0$ and $N\simeq 4.40\times 10^5$, which comes from the ground-state particle number at $\mu=(10/9)\mu_{\rm exp}$. The reason why we put a factor $10/9$ to $\mu_{\rm exp}$ is that $\mu_{\rm exp}$ is measured at the end of the experiment \cite{Campbell_pb}. According to Ref.~\cite{Kumar2017}, the atom loss reduces the chemical potential by $10\%$. Therefore, the initial particle number is evaluated by $(10/9)\mu_{\rm exp}$. 

The time dependence of the potential strength $U_0$ is determined by mimicking the experimental sequences (see Fig.~1 in Ref.~\cite{Supplemental_Kumar}). {Specifically, we start with $U_0=0$ at $t = 0$. The first $1$\;s is the preparation stage for the $W=1$ state. The strength is linearly ramped up in $300\;{\rm ms}$, kept constant ($U_0=1.1\mu_{\rm exp}$) in $400\;{\rm ms}$, and linearly ramped down in $300\;{\rm ms}$. During this ramp-up process, we rotate the barrier potential at the angular velocity $\Omega_0\equiv \hbar/(m R^2)\simeq 2\pi\times 0.88\;{\rm Hz}$. For $1\;{\rm s}\le t \le 3\;{\rm s}$, the potential strength is kept zero. Finally, $U_0$ is linearly ramped up to the desired value in $70\;{\rm ms}$ and kept constant after that.

We use the fourth-order Runge-Kutta method for the time evolution and the pseudospectral method for space discretization. The numerical meshes are used for $128\times 128$ and the time step $\Delta t$ for $\Delta t=0.5\;\mu{\rm s}$.

The time dependence of the angular momentum of the $z$ component per particle is shown in Fig.~\ref{fig:angular_momentum}, where the expression of the angular momentum is given by 
\begin{align}
L_z(t)\equiv\int d\bm{r}\Psi^{\ast}(\bm{r}, t)l_z\Psi(\bm{r},t),\label{eq:definition_of_expectation_value_of_Lz} \\
l_z\equiv -i\hbar\left(x\frac{\partial}{\partial y}-y\frac{\partial}{\partial x}\right).\label{eq:definition_of_angular_momentum_operator}
\end{align}
We can see a sudden jump of the angular momentum immediately after $t=3{\rm s}$. This is due to the ramp-up of the potential strength $U_0$ during $3{\rm s}\le t\le 3.07{\rm s}$, which does not change the winding number. After this sudden decrease, the angular momentum decays at a certain time for $U_0\ge 0.7\mu_{\rm exp}$. This decay can be attributed to the fact that the energy barrier $E_{\rm B}^{\pm}$ diminishes as the total particle number decreases via three-body losses. This result means that three-body losses can induce the decay of superflow within a few tens of seconds at zero temperature.

Here, we comment on the fluctuations of the angular momentum around zero after the phase slip. After the vortex passes through the ring trap, various oscillation modes are excited. Hence the system has weak dissipation (three-body loss), these excitations do not decay. Consequently, the angular momentum strongly fluctuates around $L_z=0$.

\subsubsection{Discussion}\label{sec:discussion_2D}

\begin{figure}[t]
\centering
\includegraphics[width=8.0cm]{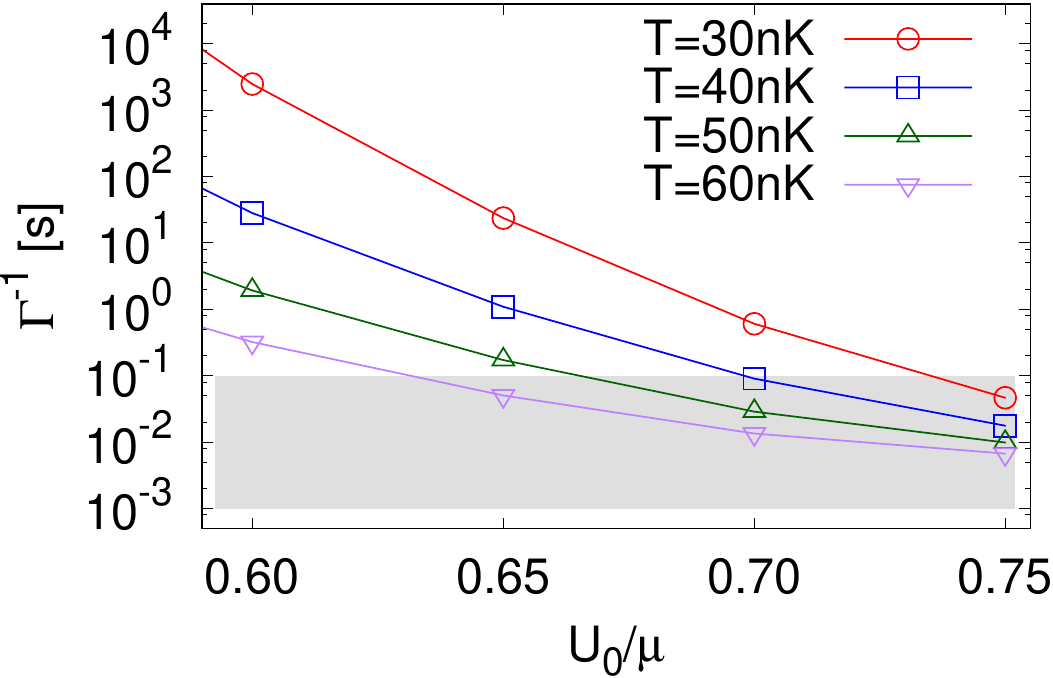}
\caption{(Color online) Lifetime of the superflow ($\Gamma^{-1}$) for small particle number $N\simeq 2.7\times 10^4$ compared to the experiment. $\mu=h\times 0.5\;{\rm kHz}$. The gray region represents $1\:{\rm ms}\le \Gamma^{-1}\le 100\;{\rm ms}.$
}
\label{fig:lifetime_small_system}
\end{figure}%

In Sec.~\ref{sec:Three-body_loss_2D}, we found that three-body losses can induce the decay of superflow at zero temperature. The remaining problem is whether this mechanism is the origin of the decay observed in the NIST experiment or not. To investigate this, we compare the decay time scale of the experiment with the theoretical one. Before discussing our results, we briefly explain how to extract the lifetime in the NIST experiment \cite{Kumar2017}. The experimentalists measured the winding number of the BEC as a function of the temperature, the barrier strength, and the time by using the interference technique \cite{Eckel2014_2}. The ensemble average of the winding number can be interpreted as the angular momentum per particle. They fitted the results to the exponential function $\exp{(-\Gamma t)}$ [see Fig.~2 (a) in Ref.~\cite{Kumar2017}]. The experimental lifetime is given by $1/\Gamma$.

The experimental results show that the lifetime is about $6.7\:{\rm s}$ for $U_0=0.6884\mu_{\rm exp}$. In the theoretical calculations, we obtain that the time for the instability to set in is about $20.4\;{\rm s}$ for $U_0=0.6884\mu_{\rm exp}$. This result shows that the theoretical decay time scale is about three times longer than the experimental one. This means that the experimental results cannot be attributed solely to the three-body losses. In addition to this, our calculations cannot explain the temperature dependence of the lifetime.

An obvious reason for this discrepancy is the lack of finite-temperature effects in our GP calculations. These effects should enhance the decay of superflow. Moreover, other mechanisms, which accelerate the decay of superflow, may exist in the actual experimental setup. For example, one-body loss due to the inelastic collision between the atoms and the background gases, inelastic collision of the light scattering, and inevitable experimental noise, and so on. Hence our results should be interpreted as the upper bound of the decay time scale. The hybrid effects of the three-body loss and the finite temperature effects are a possible scenario of the decay of superflow in the NIST experiment.

To confirm whether or not the hybrid effects appear in the experiment, we need to perform the dynamical simulations including the finite temperature effects and the three-body loss and compare them with the experiments. For example, the truncated-Wigner approximation (TWA) \cite{Blakie2008,Polkovnikov2010}, the Zaremba-Nikuni-Griffin (ZNG) formalism \cite{Griffin2009}, and the stochastic projected GP (SPGP) method \cite{Blakie2008,Rooney2013} have been used for simulating dynamics of BEC at finite temperatures in previous literature. Among them, the ZNG formalism cannot describe the nucleation process of the vortices due to thermal fluctuations. The TWA at finite temperature and the SPGP method are possible ways to attack this problem. To investigate the decay problem by using such kind of methods will be our future work.

Nevertheless, we can at least predict that, if the decay is indeed due to three-body losses, one can experimentally observe the decay via TAPS by suppressing the energy barrier in the following way: we seek the parameter region where the lifetime due to the TAPS becomes $1$-$100\;{\rm ms}$ by tuning the particle number in the trap. In this time scale, we can neglect the effects of three-body losses. Figure \ref{fig:lifetime_small_system} shows the lifetime of the superflow for $N\simeq 2.7\times 10^4$ [note that the particle number in the experiment is $O(10^5)$], which corresponds to the ground-state particle number for $\mu=h\times 0.5\;{\rm kHz}$. Here, other parameters such as trap frequencies are fixed. If these results agree with future experiments, we can conclude that the superflow decay via TAPS occurs without being affected by three-body losses.

\subsection{3D results}\label{subsec:3D}

In Sec.~\ref{subsec:Quasi-2D}, we showed the quasi-2D results. However, the confinement in the $z$ direction is not sufficiently strong for quantitatively justifying the quasi-2D approximation. In fact, the experimental parameters are given by $\omega_z=2\pi\times 974{\rm Hz}$, $\omega_{\rm r}=2\pi\times 256{\rm Hz}$, $k_{\rm B}T=k_{\rm B}\times 30{\rm nK}\simeq h\times 600{\rm Hz}$, and $\mu=h\times 2.91{\rm kHz}$. From these values, we can obtain the following inequality; $\mu> \hbar\omega_z>k_{\rm B}T>\hbar\omega_{\rm r}$, which is inconsistent with a quasi-2D condition $\hbar\omega_z\gg \max(k_{\rm B}T, \mu, \hbar\omega_{\rm r})$. In the experimental setup, the thermal excitations along the $z$ direction are suppressed and the spatial dependence of the order parameter on the $z$ direction is not of the Gaussian form. This means that the conclusions in the previous section should be treated carefully. 

In this section, we show the 3D calculation results in order to clarify the validity of the quasi-2D calculations performed in the previous section. We show the energy barrier and the real-time dynamics with the same parameters as those of the quasi-2D systems. We will see that the quasi-2D and 3D results are qualitatively similar. 

\subsubsection{Density and phase profile}\label{sec:density_and_phase_3D}

\begin{figure}[t]
\centering
\includegraphics[width=8.5cm]{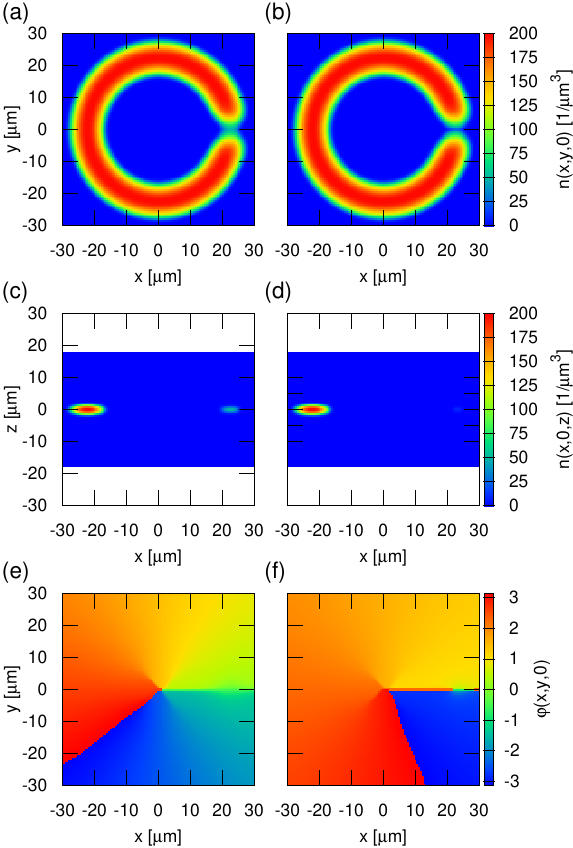}
\caption{(Color online) Density and phase profile for 3D systems. (a), (b) Density profiles in the $x$-$y$ plane ($z=0$) for metastable and unstable states. (c), (d) Density profiles in the $x$-$z$ plane ($y=0$) for metastable and unstable states. (e), (f) Phase profiles in the $x$-$y$ plane ($z=0$) for metastable and unstable states. The potential strength is $U_0=0.7\mu_{\rm exp}$.
}
\label{fig:density_and_phase_profile_3D}
\end{figure}%

First, we show the density and phase profiles in the 3D systems for the metastable and unstable stationary states in Fig.~\ref{fig:density_and_phase_profile_3D}. Here, the particle number is $N_{\rm 3D}\simeq 4.31\times 10^5$, which is evaluated as the particle number of the ground state at the chemical potential $\mu_{\rm exp}$ by solving the GP equation. We see the SV at the low-density region in the unstable solution [Figs.~\ref{fig:density_and_phase_profile_3D} (b) and \ref{fig:density_and_phase_profile_3D} (f)]. The existence of the SV reflects the $z$ dependence of the density [Fig.~\ref{fig:density_and_phase_profile_3D} (d)]. Comparing Fig.~\ref{fig:density_and_phase_profile_3D} (c) with Fig.~\ref{fig:density_and_phase_profile_3D} (d), we can see that the density around $x\sim R$ of the unstable state is lower than that of the metastable state. This is due to the existence of the SV. These results are consistent with the quasi-2D ones (see Fig.~\ref{fig:density_and_phase_profile}).

\subsubsection{Energy barrier}\label{sec:energy_barrier_3D}

\begin{figure}[t]
\centering
\includegraphics[width=8.0cm]{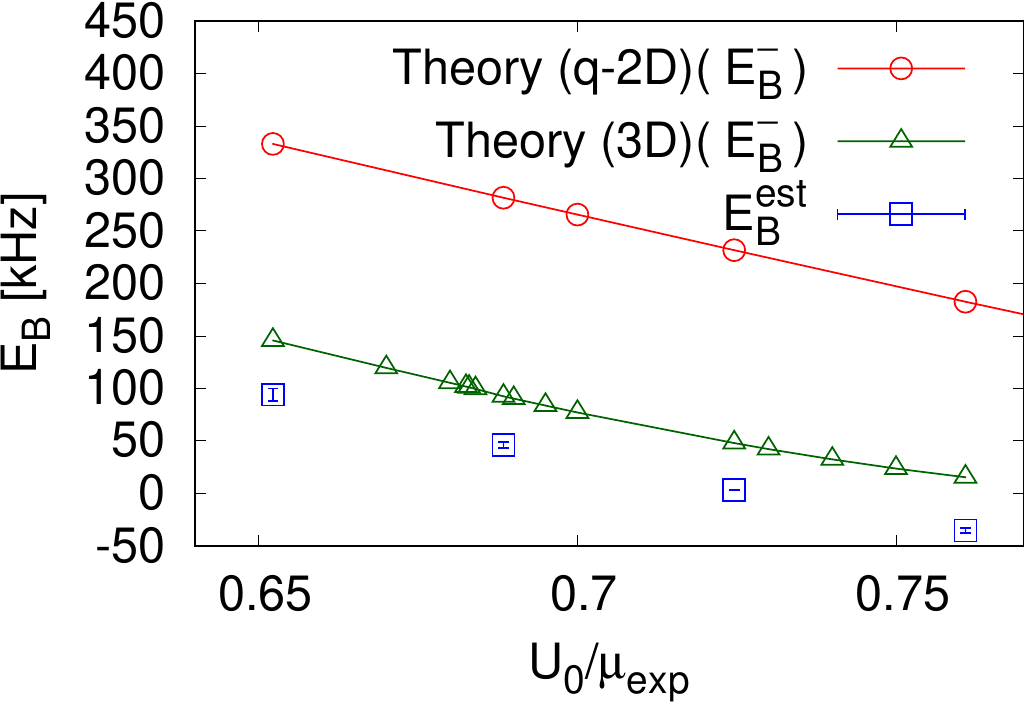}
\caption{(Color online) Quasi-2D and 3D energy barrier as a function of the strength of the external potential. The red circles and the  triangle are the value computed in the quasi-2D and 3D systems, respectively. The blue squares denote the values estimated in Ref.~\cite{Kumar2017}.
}
\label{fig:energy_barrier_3D}
\end{figure}%

\begin{figure}[t]
\centering
\includegraphics[width=8.0cm]{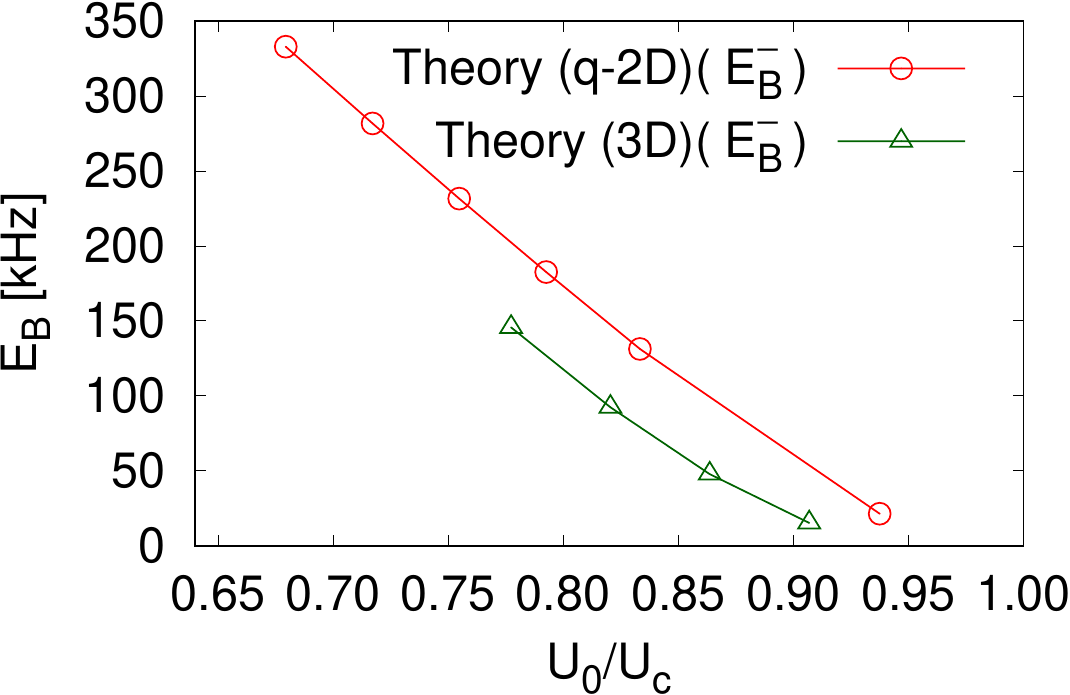}
\caption{(Color online). Energy barrier as a function of the rescaled strength of the external potential. The critical barrier strength $U_{\rm c}$ for 2D and 3D systems is given by  $U_{\rm c}^{\rm 2D}\simeq 0.960\mu_{\rm exp}$ and $U_{\rm c}^{\rm 3D}\simeq 0.839\mu_{\rm exp}$
}
\label{fig:energy_barrier_rescale}
\end{figure}%

Next, we show the energy barrier in Fig.~\ref{fig:energy_barrier_3D}. The 3D results are shown by the green triangles. We can find that the values of the energy barrier for 3D systems are close to those of the experimental one compared to the 2D ones. In the following, we explain the reason why the energy barriers in 3D are so different from that in 2D (the difference is roughly $200{\rm kHz}$).

The difference between 2D and 3D systems can be attributed to the discrepancy in the critical strength of the external potential $U_{\rm c}$, where the energy barrier for the $W=1$ states vanishes. The critical values are given by $U_{\rm c}^{\rm 2D}\simeq 0.960\mu_{\rm exp}$ and $U_{\rm c}^{\rm 3D}\simeq 0.839\mu_{\rm exp}$, which are determined by the time evolution of the GP equation as performed in Sec.~\ref{sec:Three-body_loss_2D} and Sec.~\ref{sec:Three-body_loss_3D}. This result indicates that one should compare the energy barriers in 2D and 3D systems as a function of $U_0/U_{\rm c}$ instead of $U_0/\mu_{\rm exp}$.  In Fig.~\ref{fig:energy_barrier_rescale}, we show the energy barrier as a function of rescaled external potential strength $U_0/U_{\rm c}$. This result shows that the difference between the 2D and 3D energy barrier at the same $U_0/U_{\rm c}$ is about $50{\rm kHz}$, which is small compared to the difference that is extracted from the same $U_0/\mu_{\rm exp}$.

Our results show that the values of the 3D energy barrier are small compared to those of the 2D ones. Nevertheless, our conclusion that the TAPS is irrelevant to the supercurrent decay observed in the NIST experiment still remains. The reasons are as follows. The typical energy barrier for the 3D system is given by $E_{\rm B}\sim h\times 100{\rm kHz}$. This means that the ratio between the energy barrier and the temperature is given by $E_{\rm B}/(k_{\rm B}T)\sim 100$ for $T=30{\rm nK}$. Although we do not calculate the prefactor of the Kramers formula in 3D systems due to the high computational costs, we can estimate the lifetime by using the 2D prefactor. The lifetime for $T=30{\rm nK}$ and $U_0=0.6884\mu_{\rm exp}$ becomes $\Gamma^{-1}\sim 10^{57}{\rm s}$, which is much longer than the experimental timescales. From this 3D calculation, we can confirm that the TAPS is irrelevant in the NIST experiment.

\subsubsection{Three-body loss}\label{sec:Three-body_loss_3D}

\begin{figure}[t]
\centering
\includegraphics[width=8.0cm]{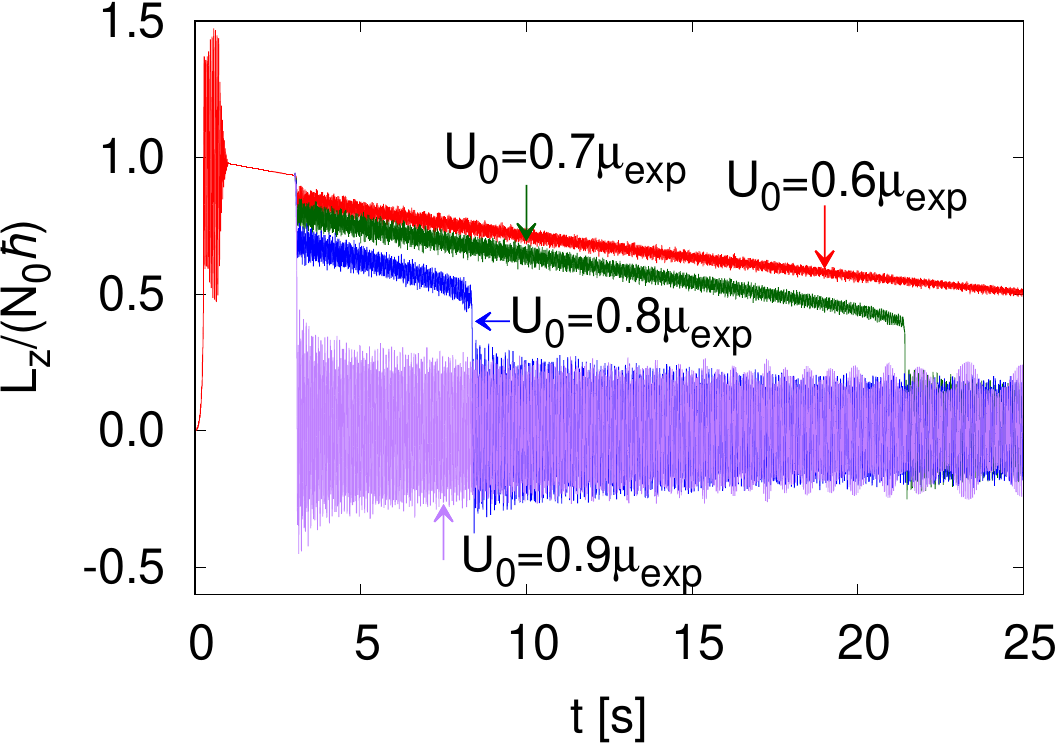}
\caption{(Color online) Time evolution of the angular momentum for various strength of the potential in 3D. The angular momentum is normalized by the initial particle number $N_0\simeq 5.32\times 10^5$.
}
\label{fig:angular_momentum_3D}
\end{figure}%

Finally, we show the results for the effects of three-body losses in the 3D systems. We solve the 3D dissipative GP equation (\ref{eq:3D_GP_equation}). The numerical procedures are the same as the quasi-2D case. The initial condition is the ground-state solution of the nondissipative GP equation ($L_3=0$) for $U_0=0$ and $N\simeq 5.32\times 10^5$, which comes from the ground-state particle number at $\mu=(10/9)\mu_{\rm exp}$. We use the split-step method for the time evolution and the pseudospectral method for space discretization. The numerical meshes and the time step used here are $128\times 128\times 64$ and $\Delta t=0.5\;\mu{\rm s}$, respectively.

Figure \ref{fig:angular_momentum_3D} shows the time evolution of the $z$ component of the angular momentum for various potential strengths. Comparing these results with those shown in Fig.~\ref{fig:angular_momentum}, the qualitative behavior is rather similar to the quasi-2D one. We also find that the time for the instability to set in is about 20.3 {\rm s} for $U_0=0.6884\mu_{\rm exp}$. This value is almost the same as  the quasi-2D case (see Sec.~\ref{sec:discussion_2D}). From these results, we conclude that the effects of the $z$ direction for the three-body loss induced decay are not significant.

\section{Summary}\label{sec:Summary}

We have investigated the supercurrent decay of BEC in ring traps to reveal the origin of the decay observed in the experiment of Ref.~\cite{Kumar2017}. First, we reconsidered the decay via the TAPS based on the Kramers formula in the quasi-2D systems. Our results show that the lifetime via TAPS is astronomically long. This means that the TAPS is not relevant in the experiment of Ref.~\cite{Kumar2017}. The same conclusion has also been obtained from the 3D calculations.

Next, we performed the numerical simulations of the GP equation with the three-body loss term in the quasi-2D and 3D systems and found that three-body losses can induce the decay of superflow at zero temperature. Comparing the decay time scale of the theory and experiment, we found the decay time scale is about three times longer than the experimental one. These results indicate that the experimental results cannot be attributed solely to the three-body loss. We proposed a possible scenario that the hybrid effects of the three-body loss and the finite temperature is the origin of the decay observed in the NIST experiment.  We also proposed that one can enhance the TAPS by decreasing the energy barrier in order to observe the decay via the TAPS in future experiments. 

We finally emphasize that the effects of particle loss, to our knowledge, have not been considered in the context of superflow decay before the present work. In this sense, our results may open up a new possibility in the study of superfluidity. Specifically, consideration of such effects is expected to be relevant also to advanced superfluid systems, including exciton polariton BEC \cite{Deng2010} and ultracold atomic gases with controllable particle losses \cite{Labouvie2016,Tomita2017}.

\begin{acknowledgments}
We thank G. K. Campbell for providing us the experimental data and useful comments, A. Kumar for providing us with the experimental data, and P. Naidon for critical reading of the manuscript. M.K. thanks S. Goto for his advice regarding numerical calculations. I.D. thanks L. Mathey for useful discussions. M.K. was supported by Grant-in-Aid for JSPS Research Fellow Grant No. JP16J07240. I.D. was supported by KAKENHI grants from JSPS: Grants No.~15H05855, No.~25220711, No.~18K03492, and No.~18H05228, by research grant from CREST, JST, and by Q-LEAP program of MEXT, Japan.
\end{acknowledgments}


\appendix
\section{DVR method}\label{sec:Appendix_DVR}

In this appendix, we explain the DVR method \cite{Baye1986}. For numerical calculations, we consider a box region $[-L_x'/2, +L_x'/2]\times[-L_y'/2, +L_y'/2]$, where $L_x'\equiv L_x/a_{\rm r}$ and $L_y'\equiv L_y/a_{\rm r}$ are the dimensionless system sizes for $x$ and $y$ directions, respectively, and $a_{\rm r}\equiv \sqrt{\hbar/(m\omega_{\rm r})}$ is the harmonic-oscillator length of the radial direction. We discretize the space into $N_x\times N_y$ lattice points. The condensate wave function is expanded as a series of the DVR functions:
\begin{align}
\Psi'(\bm{r}', t')&=\sum_{i=-(N_x-1)/2}^{(N_x-1)/2}\;\sum_{j=-(N_y-1)/2}^{(N_y-1)/2}\Psi'_{i j}(t')f_i(x')g_j(y'),\label{eq:expansion_of_two-dimensional_condensate_wave_function}
\end{align}
where $N_x$ and $N_y$ are odd natural numbers and the DVR functions $f_i(x')$ and $g_j(y')$ are defined by
\begin{align}
f_i(x')\equiv \frac{1}{N_x\sqrt{\Delta x}}\frac{\sin[\pi(x'/\Delta x -i)]}{\sin[\pi(x'/\Delta x -i)/N_x]},\label{eq:definition_of_DVR_function_x_2D}\\
g_j(y')\equiv \frac{1}{N_y\sqrt{\Delta y}}\frac{\sin[\pi(y'/\Delta y -j)]}{\sin[\pi(y'/\Delta y -j)/N_y]}.\label{eq:definition_of_DVR_function_y_2D}
\end{align}
Here, the mesh sizes of the $x$ and $y$ directions are given by $\Delta x\equiv L_x'/N_x$ and $\Delta y\equiv L_y'/N_y$. We can show that, on the mesh points, which are $x_i'\equiv \Delta x\times i$ and $y_j'\equiv \Delta y\times j$, the DVR functions satisfy $f_i(x_k)=\delta_{i,k}/\sqrt{\Delta x}$ and $g_j(y_k')=\delta_{j,k}/\sqrt{\Delta y}$. We can also show that the DVR functions satisfy the orthonormal relations
\begin{align}
&\int^{+L_x'/2}_{-L_x'/2}d x'f_i(x')f_j(x')=\delta_{i j},\label{eq:orthonormal_properties_of_DVR_function_x_2D}\\
&\int^{+L_y'/2}_{-L_y'/2}d y'g_i(y')g_j(y')=\delta_{i j},\label{eq:orthonormal_properties_of_DVR_function_y_2D}
\end{align}
where we used the Gauss quadrature to derive these relations \cite{Numerical_recipes}. We note that the wave function on the mesh points satisfies $\Psi'(x_i', y_j', t')=\Psi'_{i j}(t')/\sqrt{\Delta x\Delta y}.$

From the above relations, we can derive the matrix elements of the  first and second derivative terms
\begin{align}
T^{(1)}_{x, i j}&\equiv \int^{+L_{x}'/2}_{-L_{x}'/2}d x' f_i(x')\frac{d}{d x'}f_j(x')\notag \\
&=
\begin{cases}
\vspace{1.0em}\displaystyle{\frac{\pi}{N_{x}\Delta x}\frac{(-1)^{i-j}}{\sin[\pi(i-j)/N_x]}}\quad (i\not=j)\\
0\quad (i=j)
\end{cases}
,\label{eq:matrix_element_T1_x_2D}
\end{align}
\begin{align}
T^{(1)}_{y, i j}&\equiv \int^{+L_{y}'/2}_{-L_{y}'/2}d y' g_i(x')\frac{d}{d y'}g_j(y')\notag \\
&=
\begin{cases}
\vspace{1.0em}\displaystyle{\frac{\pi}{N_y\Delta y}\frac{(-1)^{i-j}}{\sin[\pi(i-j)/N_y]}}\quad (i\not=j)\\
0\quad (i=j)
\end{cases}
,\label{eq:matrix_element_T1_y_2D}
\end{align}
\begin{align}
T_{x, i j}^{(2)}&\equiv -\frac{1}{2}\int^{+L_x'/2}_{-L_x'/2}d x'f_i(x')\frac{d^2}{d x'^2}f_j(x')\notag \\
&=
\begin{cases}
\vspace{1.0em}\displaystyle{\frac{(-1)^{i-j}}{\Delta x^2}\frac{\pi^2}{N_x^2}\frac{\cos{[\pi(i-j)/N_x]}}{\sin^2{[\pi(i-j)/N_x]}} }\quad (i\not=j)\\
\displaystyle{\frac{\pi^2}{6\Delta x^2}\left(1-\frac{1}{N_x^2}\right) }\quad (i=j)
\end{cases}
,\label{eq:matrix_element_T2_x_2D}\\
T_{y, i j}^{(2)}&\equiv -\frac{1}{2}\int^{+L_y'/2}_{-L_y'/2}d y'g_i(y')\frac{d^2}{d y'^2}g_j(y')\notag \\
&=
\begin{cases}
\vspace{1.0em}\displaystyle{\frac{(-1)^{i-j}}{\Delta y^2}\frac{\pi^2}{N_y^2}\frac{\cos{[\pi(i-j)/N_y]}}{\sin^2{[\pi(i-j)/N_y]}} }\quad (i\not=j)\\
\displaystyle{\frac{\pi^2}{6\Delta y^2}\left(1-\frac{1}{N_y^2}\right) }\quad (i=j)
\end{cases}
.\label{eq:matrix_element_T2_y_2D}
\end{align}

One of the advantages of the DVR method is that the matrix elements of the potential and the interaction terms are diagonal. Although the matrix elements of the kinetic terms have the off-diagonal elements, the matrix is sparse. This is the advantage to the calculation of the Bogoliubov equation. The other advantage is that the DVR method has higher accuracy compared to the standard discretization method (centered difference method). The DVR method can represent up to the polynomials of degree $O(N_i)$ exactly, where $N_i (i=x,y)$ is the number of lattice points for $i$ direction. This is in contrast to the standard discretization method, which can represent up to the quadratic polynomials exactly. This fact comes from the Gauss quadrature in the derivation. 

We note that another type of basis has been used to calculate the GP and the Bogoliubov equation in Ref.~\cite{Schneider1999}.


\end{document}